\documentclass[twocolumn,showpacs,amsmath,amssymb,floatfix,prl]{revtex4}

\usepackage{graphicx} % include figure files
\usepackage{bm} % bold math
\newcommand{\rot}{{\bm\nabla}\wedge}

\begin{document}

\title{Macroscopic dynamics of a Bose-Einstein condensate containing a vortex lattice}
\author{Marco Cozzini}
\author{Sandro Stringari}
\affiliation{Dipartimento di Fisica, Universit\`a di Trento and BEC-INFM, I-38050 Povo, Italy}

\date{\today}

\begin{abstract}
Starting from the equations of rotational hydrodynamics we study the macroscopic behaviour of a trapped
Bose-Einstein condensate containing a large number of vortices. The stationary configurations of the system,
the frequencies of the collective excitations and the expansion of the condensate are investigated as a
function of the angular velocity of the vortex lattice. The time evolution of the condensate and of the
lattice geometry induced by a sudden deformation of the trap is also discussed and compared with the recent
experimental results of P.~Engels \textit{et al.}, Phys. Rev. Lett. {\bf89}, 100403 (2002).
\end{abstract}

\pacs{03.75.Fi, 32.80.Lg}

\maketitle

After their experimental realization \cite{jila1,ens1} quantized vortices have become a popular subjet of
research in the field of ultra cold gases (see for example \cite{fetter}). More recently, special emphasis has
been given to the study of configurations containing a large number of vortical lines (vortex lattices)
\cite{mit,jila2}. An important motivation is due to the possibility of reaching critical regimes of high
angular velocities where the degeneracy of the single particle levels gives rise to new quantum
configurations, analog to the quantum Hall effect \cite{wilkin,cirac,ho,macdonald}. 

The purpose of the present work is to describe the macroscopic behaviour of a Bose-Einstein condensate
containing a vortex lattice in the so called Thomas-Fermi regime \cite{review}, where the interaction energy
is much larger than the trapping oscillator energies.
A complementary investigation concerns the dynamic behaviour of vortex lines, where interactions play a role
at a more microscopic scale \cite{anglin,baym}.

In the presence of a large number of vortex lines a useful concept is the so called diffused vorticity given
by
\begin{equation} \label{eq:rot v}
\rot{\bf v} = 2\,{\bf\Omega} \ ,
\end{equation}
where ${\bf v}$ is the velocity field of the fluid and $\bf\Omega$ is assumed to be uniform. In terms of the
density $n_v$ of the vortex lines per unit surface one has $n_v=2\,\Omega\,M/h$, where $M$ is the atomic mass.
The quantity $\bf\Omega$ corresponds to the angular velocity of the sample. The concept of diffused vorticity
is adequate to describe the dynamics at macroscopic distances, larger than the average distance $n_v^{-1/2}$
between vortices.

The macroscopic description is provided by the equations of rotational hydrodynamics
\begin{equation} \label{eq:rotHD dens}
\frac{\partial{n}}{\partial t}+{\bm\nabla}\cdot({n}{\bf v}) = 0 \ ,
\end{equation}
\begin{equation} \label{eq:rotHD vel}
M\frac{\partial{\bf v}}{\partial t}+{\bm\nabla}\left(\frac{Mv^2}{2}+V_\text{ext}+g{n}\right) = 
M{\bf v}\wedge(\rot{\bf v})
\end{equation}
written in the laboratory frame, where ${n}({\bf r},t)$ is the spatial density, ${\bf v}({\bf r},t)$ is the
velocity field, $V_\text{ext}$ is the external potential and $g=4\pi\hbar^2a/M$ is the coupling costant
expressed in terms of the $s$-wave scattering length $a$.
These equations generalize the usual equations of irrotational hydrodynamics ($\rot{\bf v}=0$), which have
been successfully applied to study the collective oscillations of the condensates in the absence of vortex
lines \cite{stringari}. It is worth noticing that the hydrodynamic theory is well suited to study also non
linear effects, including the expansion of the gas after release of the confining trap. 

For an axi-symmetric harmonic potential $V_\text{ext}=M\,[\omega_\perp^2\,(x^2+y^2)+\omega_z^2\,z^2]/2$ the
equilibrium solutions of Eqs.~(\ref{eq:rotHD dens}) and (\ref{eq:rotHD vel}) have the form ${\bf
v}_0={\bf\Omega_0}\wedge{\bf r}$ and
\begin{equation} \label{eq:dens staz}
{n}_0({\bf r}) = \frac{1}{g}\,
\left\{\mu-\frac{M}{2}\,\left [\tilde\omega_\perp^2\,(x^2+y^2)+\omega_z^2\,z^2\right]\right\}\ ,
\end{equation}
where the effective frequency given by $\tilde\omega_\perp^2=\omega_\perp^2-\Omega_0^2$ originates from the
centrifugal effect and
\begin{equation} \label{eq:mu}
\mu = \mu_0\,\left[1-\left(\frac{\Omega_0}{\omega_\perp}\right)^2\right]^{2/5}
\end{equation}
is the chemical potential. In the previous equation
$\mu_0=(\hbar\omega_\text{ho}/2)\,(15Na/a_\text{ho})^{2/5}$ is the chemical potential in the absence of
rotation, $\omega_\text{ho}=(\omega_\perp^2\omega_z)^{1/3}$ is the average oscillator frequency,
$a_\text{ho}=\sqrt{\hbar/M\omega_\text{ho}}$ is the corresponding oscillator length and $N$ is number of
atoms. These stationary solutions are defined only for $\Omega_0<\omega_\perp$. The density profile
(\ref{eq:dens staz}) exhibits a typical bulge effect produced by the centrifugal force, giving rise to the
angular velocity dependence
\begin{equation} \label{eq:asp ratio}
\frac{R_\perp}{R_z} = \frac{\omega_z}{\sqrt{\omega_\perp^2-\Omega_0^2}}
\end{equation}
for the aspect ratio, where $R_\perp$ and $R_z$ are, respectively, the Thomas-Fermi radii of the atomic cloud
in the radial and axial directions.
Result (\ref{eq:asp ratio}) can be used to deduce the value of $\Omega_0$ from the \textit{in situ}
measurement of the aspect ratio. In Ref.~\cite{jila2} values up to $\Omega_0/\omega_{\perp}=0.95$ have been
obtained. In that experiment the condensate takes a pancake form even if the confining trap has a cigar shape
($\omega_z<\omega_\perp$).

The angular momentum per particle carried by the system is given by 
\begin{equation} \label{eq:ang mom}
\langle l_z \rangle = M\Omega_0\langle x^2+y^2 \rangle =
\frac{4}{7}\,\frac{\Omega_0\ }{\omega_\perp^2-\Omega_0^2}\,\mu
\end{equation}
and becomes larger and larger as $\Omega_0\to\omega_\perp$.

The collective oscillations of the condensate are obtained by looking for the linearized solutions of
Eqs.~(\ref{eq:rotHD dens}) and (\ref{eq:rotHD vel}) in the form ${n}={n}_0+\delta{n}$ and ${\bf v}={\bf
v}_0+\delta{\bf v}$ with $\delta{n}$ and $\delta{\bf v} \sim e^{-i\omega t}$.

If one chooses $\delta{n}=a_\pm(x\pm iy)^2$, $\delta{\bf v}=\alpha_\pm{\bm\nabla}(x\pm iy)^2$, corresponding
to quadrupole excitations with $m=\pm2$ where $m$ is the third component of angular momentum, one easily finds
the dispersion law
\footnote{Eq.~(\ref{eq:m+-2}) can be generalized to excitations of higher multipolarity $\ell$ and
$m=\pm\ell$. One finds \cite{chevy}
$\omega_\pm=\sqrt{\ell\,\omega_\perp^2-(\ell-1)\,\Omega_0^2}\pm(\ell-1)\,\Omega_0$.}
\begin{equation} \label{eq:m+-2}
\omega(m=\pm2) = \sqrt{2\,\omega_\perp^2-\Omega_0^2}\pm\Omega_0 \ .
\end{equation}
These frequencies have been directly measured in the experiment of \cite{jila2}, confirming with high accuracy
the predictions of theory. The splitting $2\,\Omega_0$ between the two frequencies agrees with the sum rule
result \cite{zambelli}
\begin{equation}
\omega(m=+2)-\omega(m=-2) = \frac{2}{M}\,\frac{\langle l_z \rangle}{\langle x^2+y^2 \rangle} \ ,
\end{equation}
as can be seen using the rigid body expression (\ref{eq:ang mom}) for the angular momentum.

When $\Omega_0\to\omega_\perp$ one finds $\omega(m=+2)\to2\omega_\perp$, while $\omega(m=-2)\to0$, reflecting
the tendency of the system to become unstable against static quadrupole deformations.

In addition to the $m=\pm2$ oscillations, it is interesting to discuss the behaviour of the $m=0$ modes
\footnote{The $m=\pm1$ modes, wich involve a dynamic coupling with the rotation of the axis of vorticity
(tilting mode), will be discussed elsewhere \cite{chevy}.}.
These result from the coupling between the axial and radial motion of the condensate. By looking for solutions
of the form $\delta{n}({\bf r})=a_0+a_\perp(x^2+y^2)+a_z\,z^2$ and $\delta{\bf v}({\bf
r})=\delta{\bf\Omega}\wedge{\bf r}+{\bm\nabla}[\alpha_\perp(x^2+y^2)+\alpha_z\,z^2]$ one finds that the two
decoupled frequencies are given by
\begin{eqnarray}
\lefteqn{\omega^2(m=0) \,\ =}\\
&&\nonumber\textstyle2\,\omega_\perp^2+\frac{3}{2}\,\omega_z^2 \pm\frac{1}{2}
\sqrt{16\,\omega_\perp^4+9\,\omega_z^4-16\,\omega_z^2\,\omega_\perp^2-8\,\omega_z^2\,\Omega_0^2} \ .
\end{eqnarray}
When $\Omega_0=0$ one recovers the solutions of \cite{stringari}, while for $\Omega_0\to\omega_\perp$ the two
frequencies approach the value $2\,\omega_\perp$ (radial compressional mode) and $\sqrt3\,\omega_z$ (axial
compressional mode). The latter value coincides with the frequency predicted by theory for a pancake geometry
($\omega_z\gg\omega_\perp$) in the absence of vortices.

The tendency of the system to become unstable against a quadrupole deformation when $\Omega_0\to\omega_\perp$
has been used in the recent experiment of \cite{jila3} to induce large deformations in the condensate and to
explore the consequences on the geometry of the vortex lattice. 

\begin{figure}

\includegraphics[width=8.5cm]{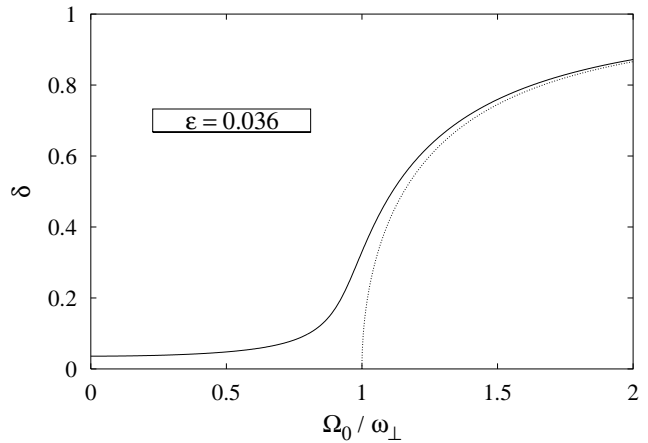}

\caption{\label{fig:delta}Deformation $\delta=\alpha/\Omega_0$ of the cloud for the steady-state solutions of
Eqs. (\ref{eq:rotHD dens}) and (\ref{eq:rotHD vel}) as a function of the angular velocity $\Omega_0$ of the
vortex lattice, for a fixed value $\epsilon=0.036$ of the trap deformation. The dotted line is the asymptotic
solution for $\epsilon\to0$.}

\end{figure}

In the presence of a static deformation of the trap
\begin{equation} \label{eq:static perturb}
\delta V_\text{ext} = \frac{M}{2}\,\omega_{\perp}^2\epsilon(x^2-y^2)
\end{equation}
the stationary solutions of the hydrodynamic equations (\ref{eq:rotHD dens}) and (\ref{eq:rotHD vel}) in the
laboratory frame change in a quite interesting  way. On the one hand the velocity field takes the form 
\begin{equation} \label{eq:vel staz def}
{\bf v}_0({\bf r}) =  {\bf\Omega_0}\wedge{\bf r}+\alpha{\bm\nabla}(xy) \ ,
\end{equation}
containing a crucial irrotational component fixed by the parameter $\alpha$. On the other hand the
equilibrium profile assumes the deformed shape 
\begin{equation} \label{eq:dens staz def}
{n}_0({\bf r}) = \frac{1}{g}\,
\left[\mu-\frac{M}{2}\,\left(\tilde\omega_x^2\,x^2+\tilde\omega_y^2\,y^2+\omega_z^2\,z^2\right)\right]\ ,
\end{equation}
with the effective frequencies defined by $\tilde\omega_{x,y}^2=\omega_{x,y}^2+\alpha^2-\Omega_0^2$ and the
chemical potential given by $\mu=\mu_0(\tilde\omega_x\tilde\omega_y/\omega_x\omega_y)^{2/5}$. Use of the
equation of continuity yields the following third order equation for $\alpha$
\begin{equation} \label{eq:alpha}
\alpha^3+\alpha(\omega_\perp^2-\Omega_0^2)-\epsilon\omega_\perp^2\Omega_0 = 0 \ ,
\end{equation}
whose solution, expressed in terms of the deformation of the condensate
\begin{equation} \label{eq:delta}
\delta = \frac{\langle y^2-x^2 \rangle}{\langle x^2+y^2 \rangle} = \frac{\alpha}{\Omega_0} \ ,
\end{equation}
is reported in Fig.~\ref{fig:delta} for the value $\epsilon=0.036$. The other solutions of the third order
equation (\ref{eq:alpha}) should be excluded because they correspond to negative values of
$\tilde\omega_{x,y}^2$. This differs from the situation of Ref.~\cite{recati}, where only irrotational flow
was considered and more stationary solutions were found to be available in the rotating frame. For
$\epsilon\ne0$, Fig.~\ref{fig:delta} shows that, differently from the axisymmetric case, a deformed system can
support values of $\Omega_0$ larger than $\omega_\perp$, the irrotational term of Eq.~(\ref{eq:vel staz def})
providing the crucial compensation to the centrifugal effect generated by the rotational component. When
$\epsilon\to0$ and $\Omega_0\geq\omega_\perp$ one finds $\tilde\omega_{x,y}^2\to0$ and the branch approaches
the asymptotic solution $\delta=\sqrt{1-(\omega_\perp/\Omega_0)^2}$.

\begin{figure}

\includegraphics[width=8.5cm]{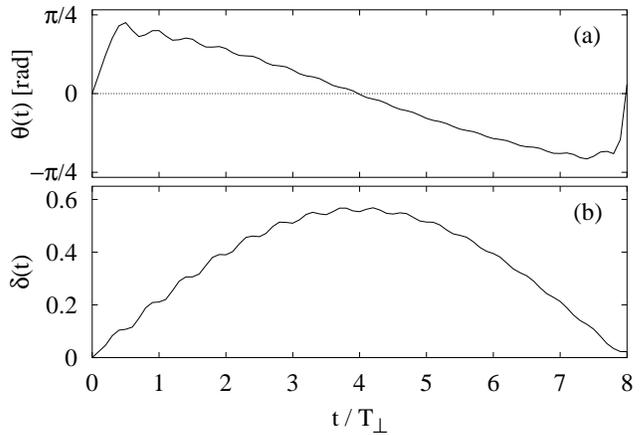}

\caption{\label{fig:ang and def}Time dependence of the condensate after switching on the trap deformation with
$\epsilon=0.036$: (a) evolution of the angle $\theta(t)$ between the principal axis of the cloud and of the
trap in the $x\,-\,y$ plane; (b) evolution of the cloud deformation $\delta(t)$. The axial frequency of the
trap is $\omega_z=0.65\,\omega_\perp$, while the initial configuration is the stationary solution for
$\epsilon=0$ and $\Omega=0.95\,\omega_\perp$. Time is measured in units of $T_\perp=2\pi/\omega_\perp$.}

\end{figure}

In the experiment of \cite{jila3} the static deformation (\ref{eq:static perturb}) was switched on suddenly at
some initial time $t=0$. This produces a time tependent perturbation which drives the system far from the
initial axi-symmetric configuration. In the following we will discuss the prediction of Eqs.~(\ref{eq:rotHD
dens}) and (\ref{eq:rotHD vel}) for the time evolution of the condensate shape as well as of the vortex lines
which are assumed to follow the motion of the fluid. The equations are easily solved in the non linear regime
by looking for the ansatz
\begin{eqnarray}
\label{eq:rho(t)}{n} & = & a_0+a_xx^2+a_yy^2+a_zz^2+a_{xy}xy \\ 
\label{eq:v(t)}{\bf v} & = &
{\bf\Omega}\wedge{\bf r}+{\bm\nabla}(\alpha_xx^2+\alpha_yy^2+\alpha_zz^2+\alpha_{xy}xy),
\end{eqnarray}
where now $\Omega$, as well as the coefficients $a_i$ and $\alpha_i$, is time dependent. In Fig.~\ref{fig:ang
and def}(a) we report the predictions for the angle between the principal axis of the condensate and the $y$
axis of the trap, with the initial value of the angular velocity $\Omega_0=0.95\,\omega_{\perp}$. The angle
first increases fast following the direction of the vortex flow. This happens because for short times the
perturbation excites with equal strength both the $m=+2$ and $m=-2$ modes and hence the system exhibits a
precession with angular velocity proportional to $\omega(m=+2)-\omega(m=-2)$. For longer times the response is
governed by the static effect and is dominated by the $m=-2$ mode, producing a precession in the opposite
direction whose period becomes longer and longer as the value of $\epsilon$ decreases. The deformation of the
condensate as a function of time is plotted in Fig.~\ref{fig:ang and def}(b) for the same conditions: rather
large values are reached, in agreement with the experimental findings of \cite{jila3}. For example at
$t\simeq2.5\,T_\perp$ the authors of \cite{jila3} report the value $\delta\simeq0.4$. The coupling with the
fast $m=+2$ mode is evident in both figures and results in the small oscillations of higher frequency.

\begin{figure}
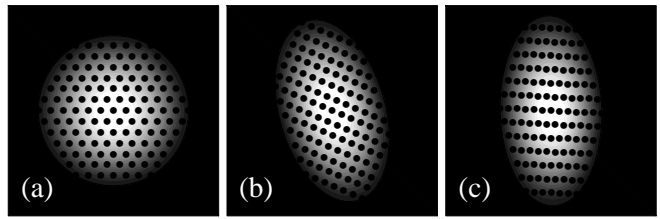


\includegraphics[width=2.8cm]{fig3a.eps}
\includegraphics[width=2.8cm]{fig3b.eps}
\includegraphics[width=2.8cm]{fig3c.eps}

\caption{\label{fig:vl-evolution}Evolution of the vortex lattice after switching on the trap deformation
$\epsilon=0.036$: (a) $t=0$, (b) $t=2.8\,T_\perp$, (c) $t=3.9\,T_\perp$.
The peculiar geometry shown in (c) appears when the cloud deformation $\delta$ is large and one of the lattice
vector is aligned along the short axis of the condensate. The trap parameters are the same as in
Fig.~\ref{fig:ang and def}.}

\end{figure}

Once the solutions of the hydrodynamic equations are known one can also study the time evolution of the vortex
lines wich are assumed to move following the velocity field ${\bf v}$ (Fig.~\ref{fig:vl-evolution}). We have
initially assumed [Fig.~\ref{fig:vl-evolution}(a)] an hexagonal geometry (Abrikosov lattice \cite{abrikosov}).
In the absence of deformation the vortex lines rotate rigidly with angular velocity $\Omega_0$. However, once
the deformation of the condensate takes place, the vortex lattice changes its geometry since it is forced to
follow the stream lines of the velocity flow (\ref{eq:v(t)}) wich contains an irrotational component. For
example, in Fig.~\ref{fig:vl-evolution}(b) one sees the formation of a near orthorhombic lattice, while for
large deformations one observes peculiar structures characterized by the compression of vortex lines along the
short radius [Fig.~\ref{fig:vl-evolution}(c)]. In our description the vortex lines are just points of the
fluid moving with velocity ${\bf v}$, so we do not account for the  possible deformation of the density
profiles induced by the vicinity of the vortex lines. These effects are expected to give rise to a significant
suppression of the density in the region between two vortices when their relative distance becomes small, with
the appearence of stripe like configurations as observed in the experiments of \cite{jila3} and recently
discussed in \cite{mueller}.

\begin{figure}

\includegraphics[width=8.5cm]{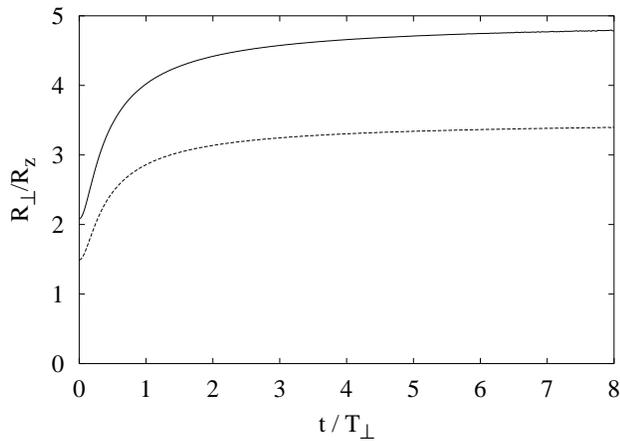}

\caption{\label{fig:expansion}Time evolution of the aspect ratio (\ref{eq:asp ratio}) after releasing the
axisymmetric trap for a condensate rotating at $\Omega_0=0.95\,\omega_\perp$ (solid line) and
$\Omega_0=0.90\,\omega_\perp$ (dashed line). The anisotropy of the trap is
$\lambda=\omega_z/\omega_\perp=0.65$.}

\end{figure}

An important role in the experimental detection of vortices is played by the expansion of the condensate. The
vortex cores are indeed too small to be imaged \textit{in situ}. It is therefore important to look at the
behaviour of the condensate after the sudden release of the trap.
This can be investigated by solving the hydrodynamic equations (\ref{eq:rotHD dens}) and (\ref{eq:rotHD vel})
setting $V_\text{ext}=0$ at the release time. The general solution is still given by the form
(\ref{eq:rho(t)}) and (\ref{eq:v(t)}).
We have first studied the time dependence of the aspect ratio (\ref{eq:asp ratio}), starting from an
axisymmetric configuration ($\epsilon=0$). Using the trapping parameters of Ref.~\cite{jila3} we find that
$R_\perp/R_z$ increases in time (see Fig.~\ref{fig:expansion}). This dramatically differs from the expansion
of a non rotating condensate, which would instead transform a pancake cloud into a cigar shape one. This
behaviour is the consequence of the large radial rotational kinetic energy, which produces a fast expansion in
the radial direction, and becomes more and more pronounced as $\Omega_0\to\omega_\perp$.
We have also investigated the time evolution of the cloud deformation $\delta(t)$ in the $x\,-\,y$ plane, by
switching off the confining potential when the perturbation (\ref{eq:static perturb}) has produced a sizable
condensate deformation [see Fig.~\ref{fig:ang and def}(b)]. Using the experimetal parameters of \cite{jila3}
we find that the value of the condensate deformation remains almost constant during the expansion. This is
again the consequence of the centrifugal effect, which is stronger in the direction of the long axis and
compensates the effects of the pressure gradient.

Let us finally discuss the conditions of applicability of the Thomas-Fermi approximation used in the present
work. If $\Omega_0$ becomes too close to $\omega_\perp$ the condition $\mu\gg\hbar\omega_\perp,\hbar\omega_z$
required to apply the Thomas-Fermi approximation is no longer valid [see Eq.~(\ref{eq:mu})].
In particular if $\mu\simeq\hbar\omega_\perp$ the size of the vortex core, fixed by the healing length
$\xi=\hbar/\sqrt{2M\mu}$, becomes comparable to the average distance between vortices, fixed by the so
called effective magnetic length $\ell_{\Omega_0}=\sqrt{\hbar/2M\Omega_0}$. In fact one has
$\xi/\ell_{\Omega_0}=\sqrt{\hbar\Omega_0/\mu}\simeq\sqrt{\hbar\omega_\perp/\mu}$.
In the experimental conditions of Ref.~\cite{jila3} one has $\mu_0\simeq27\,\hbar\omega_\text{ho}$ and the
Thomas-Fermi approximation is well satisfied also for $\Omega_0=0.95\,\omega_\perp$. Deviations from the
Thomas-Fermi regime require smaller values of $\mu_0$ and/or larger values of $\Omega_0/\omega_\perp$ and
should show up in a different behaviour of the collecitve frequencies.

Let us also mention that in order to observe quantum Hall features one should reach even more extreme
conditions, where the filling factor $\nu=N/N_v$, given by the ratio between the number of atoms and the
number of vortices, is of the order of unity
\footnote{For example, assuming a 3-D Thomas-Fermi profile, one finds $\nu=(2/15)(\mu/\hbar\Omega_0)(R_z/a)$
and hence $\nu\gg1$ even if $\mu\simeq\hbar\Omega_0$.}.
Under these conditions the low frequency oscillations in the rotating frame are expected to exhibit a chiral
behaviour \cite{cazalilla}, similar to what happens to the edge excitations in the quantum Hall effect.

In conclusion, we have studied the stationary configurations and the dynamics of a Bose-Einstein condensate
both in an axisymmetric and in an anisotropic static trap, using the equations of rotational hydrodynamic and
the concept of diffused vorticity. Special emphasis has been given to the collective oscillations, wich turn
out to be significantly affected by the rotation of the cloud. In addition, our macroscopic description gives
valuable insight on the evolution of the vortex lines recently observed in \cite{jila3}.

\begin{acknowledgments}

Useful discussions with F.~Chevy, E.~A.~Cornell and P.~Engels are acknowledged. This work was supported by the
Ministero dell'Uni\-ver\-sit\`a e della Ricerca Scientifica e Tecnologica (MURST).

\end{acknowledgments}

\end{document}